\documentclass{aa}
\usepackage{graphics}
\usepackage{times}

\newcommand{\Teff}{\mbox{$T_{\rm eff}$}}
  \newcommand\loge{$ \log{\epsilon}$}
  
  \newcommand\loghe{$ \log{\frac{n_{\rm He}}{n_{\rm H}}}$}
  \newcommand{\logg}{\mbox{\,log $g$}}                   

\newcommand{\Msol}{\mbox{$M_{\odot}$}}
\newcommand{\Rsol}{\mbox{$R_{\odot}$}}
\newcommand{\fbsdrei}{{\it Proceedings of the
Third Conference on Faint Blue Stars, {\rm eds. A.G.D. Philip,  
J. Liebert and R.A. Saffer, Schenectady: L.Davis Press}}}

\begin{document}

\sloppy
  
\thesaurus{08.01.1, 08.01.3, 08.08.2, 08.09.2, 08.15.1, 08.18.1}

\title{Spectral analysis of multi mode pulsating sdB stars\\
II. Feige\,48, KPD\,2109+4401 and PG\,1219+534
\thanks{Based on observations 
    obtained 
    at the W.M. Keck Observatory, which is operated by the Californian 
    Association for Research in Astronomy for the California Institute of 
    Technology and the University of California
}}
\author{U. Heber\inst{1}, I.N. Reid\inst{2}, K. Werner\inst{3}
}

\offprints{U. Heber}

\institute{Dr.-Remeis-Sternwarte, Universit\"at Erlangen-N\"urnberg,
           Sternwartstr. 7,
           D-96049 Bamberg, Germany
           INTERNET: heber@sternwarte.uni-erlangen.de
           \and
           Department of Physics and Astronomy, University of Pennsylvania, 
           Philadelphia, PA 19104-6396, USA
           INTERNET: inr@herschel.physics.upenn.edu
           \and
           Institut f\"ur Astronomie und Astrophysik, Universit\"at T\"ubingen,
           D-72076 T\"ubingen, Germany
           INTERNET: werner@astro.uni-tuebingen.de
}

\date{received; accepted}
\authorrunning{Heber et al.}
\titlerunning{Pulsating sdB stars}
\maketitle

\begin{abstract}
Three members of the new class of pulsating sdB stars (sdBV or EC\,14026 
stars) are analysed
from Keck HIRES spectra using line blanketed 
NLTE and LTE model atmospheres.  
Atmospheric parameters (\Teff, log~g, log(He/H)), metal 
abundances and rotational velocities
are determined. 
A careful investigation of several temperature indicators, i.e. line profile 
fitting of Balmer and helium lines, the ionization equilibria of helium, 
nitrogen and silicon gave consistent results for Feige\,48 and KPD2109$+$4401 
to within a few hundred Kelvin. 
However, for PG\,1219+534 considerably higher 
effective temperature estimates were derived from the ionization equilibria of 
nitrogen (36\,800K) and helium (34\,400\,K) than from the Balmer line profile 
fitting (33\,200\,K). A systematic difference in the gravity 
derived from NLTE and LTE models was observed, the NLTE gravities 
being slightly lower, by up to 0.1\,dex, than the LTE results.

As is typical for sdB stars,
all programme stars are found to be helium deficient, 
with a helium abundance ranging from 
1/80 solar for Feige\,48 to 1/3 solar for PG\,1219+534,
probably due to diffusion. Most metals are also depleted.
The abundances of 
C, O, Ne, Mg, Al and Si in the high gravity programme stars 
KPD\,2109+4401 and PG\,1219+534 are considerably lower than in the 
lower gravity stars  Feige\,48 and PG\,1605+072 (Heber et al., 1999)
which could be explained by an equilibrium between gravitational settling 
and radiative levitation. 
Surprisingly iron is solar to within error limits in all programme stars
irrespective of their gravity, 
confirming predictions from diffusion calculations of Charpinet et al. (1997). 

The metal lines are very sharp and allow the microturbulent velocity to be 
constrained to be lower than 5\,km/s (KPD\,2109+4401, PG\,1219+534). 
Also the projected rotational velocities have to be very low 
(v$_{rot}\,$sini $<$ 10$\,$km/s). For 
Feige\,48 the limits are even tighter (v$_{micro}\le$3\,km/s, 
v$_{rot}\,$sini $\le$ 5$\,$km/s).

\end{abstract}

\keywords{stars: atmospheres -- stars: abundances -- stars: subdwarfs -- 
          Stars: rotation -- stars: oscillations --
          stars: individual: Feige\,48, KPD\,2109$+$4401, PG\,1219$+$534
}

\section{Introduction}
It is now well established that the hot subluminous B stars can be 
identified with models of the extreme Horizontal Branch (EHB, Heber, 
1986, Saffer et al. (1994). Their photospheric chemical composition is 
governed by diffusion processes leading to strong peculiarities (e.g. 
deficiencies of helium, carbon and silicon in some stars) with strong 
variations from star to star (for a review see Heber, 1998). 

Recently, 
several sdB stars have been found to be pulsating (termed EC14026 stars 
after the prototype, see O'Donoghue et al. 1999 
for a review), 
defining a new instability strip in the HR-diagram. 
The study of these pulsators 
offers the possibility of applying  
the tools of asteroseismology to investigate the structure of sdB stars.
The existence of pulsating sdB stars was predicted by 
Charpinet et al. (1996), who uncovered an efficient driving mechanism due 
to an opacity bump associated with iron ionization in EHB models. However,
in order to drive the pulsations, iron needed to be enhanced in the appropriate 
subphotospheric layers, possibly due to diffusion. Subsequently, 
Charpinet et al. (1997) confirmed this assumption by detailed
diffusion calculations. Even more 
encouraging was the agreement of the observed and predicted instability 
strip. 

Eighteen pulsating sdB stars are well-studied photometrically (see O'Donoghue 
et al. 1999 for a review, and Bill\`eres et al. 2000, Silvotti et al. 2000 
and {\O}stensen et al. 2000 for more recent discoveries).
A precise knowledge of effective temperature, gravity, element 
abundances and rotation is
a prerequisite for the asteroseismological investigation.

\begin{figure}
\vspace{9.0cm}
\includegraphics{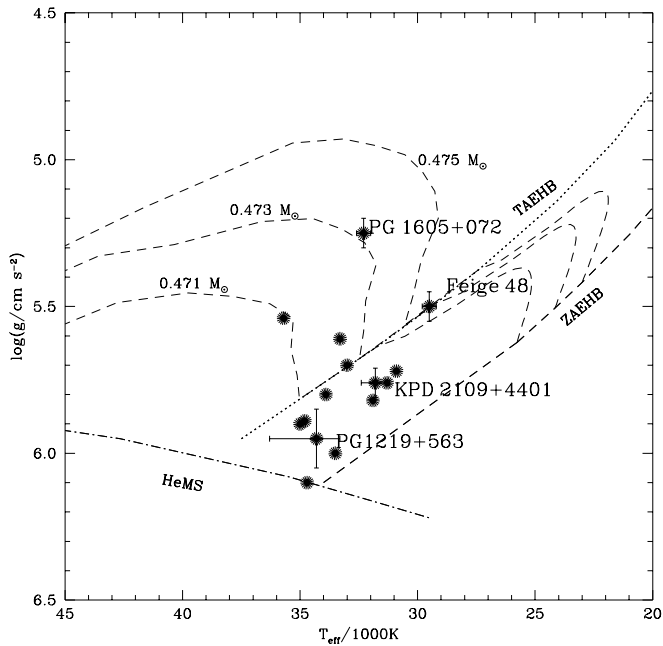}
\caption[]{Position of the programme stars in the (\Teff, log\,g) plane and 
comparison with other EC\,14024 stars which are well-studied photometrically.
The position of the zero age 
extreme Horizontal Branch (ZAEHB), the terminal age extreme Horizontal Branch
(TAEHB), the helium main sequence and evolutionary tracks for EHB stars
(Dorman et al., 1995) are also shown. \label{pulshrd}}
\end{figure}

We selected four EC14026 stars for a detailed quantitative spectral analysis
(see Fig.~\ref{pulshrd}):
PG$\,$1605$+$072 was chosen 
because it has the lowest gravity and, therefore, has probably already 
evolved beyond the extreme horizontal branch phase. It also  
displays the richest 
frequency spectrum amongst the EC$\,$14026 stars ($>$50 periods have been 
identified, Kilkenny al. 1999). The results of an spectral analysis have
already been reported by Heber et al. (1999).
Recently, Kawaler (1999) predicted from his 
modelling of the pulsations that PG$\,$1605$+$072 should be rotating. This 
prediction was confirmed by the analysis of optical spectra from which a 
projected rotational velocity of v$_{rot}$\,sin\,i=39km/s was derived 
(Heber et al., 1999).
In this paper we present the spectral analysis of three additional 
pulsating sdB stars.
PG$\,$1219$+$534 was chosen 
because it has the shortest pulsation periods and has a helium 
abundance larger than most other sdB stars (O'Donoghue et al., 1999).
Feige$\,$48 was selected because it is the coolest of all EC$\,$14026 stars 
known (Koen et al., 1998). The effective temperature and gravity of 
KPD\,2109+4401 (Bill\`eres et al., 1998) 
places the star in the parameter space where most EC\,14026 stars are 
found (see Fig.~1). 
For Feige$\,$48 and KPD$\,$2109$+$4401 only few frequencies (four for the 
former, five for the latter) have been found so far. 

\section{Observations}

High resolution optical spectra
were obtained with the 
HIRES echelle spectrograph (Vogt et al. 1994) on the Keck 
I telescope on July 20, 1998 using the blue cross 
disperser to cover the full wavelength region between 3700\AA\ and 5200\AA\
at a resolution 0.09\AA.

The spectra are integrated over one 
pulsation cycle or more since the exposure times (600--900s) were long 
compared to the pulsational periods.

The standard data reduction as described by Zuckerman \& Reid (1998)
resulted in spectral orders that have a somewhat wavy 
continuum. In order to remove the waviness we used the spectrum of H1504$+$65 
(a very hot pre-white dwarf devoid of hydrogen and helium, Werner 1991) 
which was observed in the same night. Its spectrum has only few weak lines 
of highly ionized metals in the blue (3600--4480\AA) where the strong Balmer 
lines are found in the sdB stars. Therefore we normalized 
individual spectral orders 1 to 20 (3600--4480\AA) of the sdB stars by
dividing through the smoothed spectrum of H1504$+$65. The remaining 
orders were normalized by fitting the continuum with spline functions
(interpolated for orders 26 and 27 which contain H$\beta$). 
Judged from the match of line profiles in 
the overlapping parts of neighboring orders this procedure worked 
extremely well. Atmospheric parameters determined from individual
Balmer lines are found to be consistent with each other except for H$\beta$.
Since this might be a result of interpolation errors in the normalization 
process, we excluded H$\beta$ from the fit procedure. 

Besides the strong Balmer lines of hydrogen, helium lines are present in 
all three stars. However, \ion{He}{II} 4686\AA\ is detected in KPD\,2109+4401 
and PG\,1219+534, only.
In addition, weak photospheric metal lines can be identified in the spectra of 
all programme stars. 
However, the number of detectable photospheric lines differs considerably. 
The largest number of metal lines is present in Feige$\,$48 (C, N, O, Ne, Mg, 
Si, Al, S and Fe). In PG$\,$1219$+$534 and KPD\,2109+4401 
only N, S and Fe are detectable. 
Also interstellar \ion{Ca}{II} lines are present and 
are found to have a complex structure (three to four components, see 
appendix).

\section{Spectral analysis}

The simultaneous fitting of Balmer line profiles by a grid of 
synthetic spectra has become the standard 
technique to determine the atmospheric parameters of hot high gravity stars 
(Bergeron et al. 1992). The procedure has been extended to include helium 
line profiles as well and applied successfully to sdB stars by Saffer 
et al. 1994). It will be referred to as Saffer's procedure 
throughout this paper.
The Balmer lines (H$\gamma$ to H$\,$12), He~I 
(4471\AA, 4026\AA, 4922\AA, 
4713\AA, 5016\AA, 5048\AA) and He~II 4686\AA\ lines are fitted to 
derive all three parameters simultaneously.

The analysis is based on an 
updated version of the LTE model atmosphere code of Heber et al. (1984a)
which includes metal line blanketing using Kurucz'
ATLAS6 Opacity Distribution Functions. A large grid of models is calculated
for various helium abundances and solar metalicity as well as for metal 
poor ([M/H]=-2.0) composition (see Heber et al. 1999b). 
Lemke's version\footnote{For a description see
http://a400.sternwarte.uni-erlangen.de/$\sim$ai26/linfit/linfor.html} of
the LINFOR program (developed originally by Holweger, Steffen, and
Steenbock at Kiel University) is used to compute a grid of theoretical spectra
which include the Balmer lines H$_\alpha$ to H$_{22}$ and \ion{He}{I} and 
\ion{He}{II} lines.
As can be seen from Tables \ref{res_f48}, \ref{res_kpd} and \ref{res_pg1219}
the metalicity dependence of the results is marginal, the effective 
temperature changes by 600\,K, the gravity by 0.02 dex and the helium 
abundance by 0.09 dex at the most when we apply Saffer's procedure with the 
metal poor LTE grid ([M/H]=-2.0) instead of the solar composition grid.
Tests using models calculated with Kurucz' ATLAS9 code were also performed. 
No significant deviations of the fit results from those listed in Table~1
were found.

In order to investigate the role of NLTE effects we repeated the
analyses using 
a grid of H-He line blanketed, metal free NLTE 
model atmospheres (Napiwotzki 1997), calculated with the ALI 
code of Werner \& Dreizler (1999). 
Applying Saffer's procedure with the NLTE model grid (see 
Tables~\ref{res_f48}, \ref{res_kpd} and \ref{res_pg1219}) 
yields \Teff\ and \loghe\ almost
identical to that obtained from the LTE grid. 
We therefore conclude that 
the Balmer and helium lines are not vulnerable to NLTE effects.

However, a systematic difference in log\,g persists, the LTE values
being higher by 0.1 dex than the NLTE results 
(see Table~\ref{res_f48}).
Since its origin is obscure, we finally adopted the 
averaged atmospheric parameters given in Tables~\ref{res_f48},
\ref{res_kpd} and \ref{res_pg1219}. 

The metal lines are sufficiently isolated to derive 
abundances from their equivalent widths again using the LINFOR program.
Oscillator strengths were taken from the critical compilation of 
Wiese et al. (1996) for C, N and O, from the Opacity Project (Seaton, 1987)
for \ion{Ne}{II}
\ion{Mg}{II}, and \ion{Al}{III} using the Opacity Project data base 
``TOPbase'' (Cunto \& Mendoza, 1992) at the CDS (Strasbourg),
from Becker \& Butler (1990) for Si, Wiese et al. (1969) 
for \ion{S}{III}, and Ekberg (1993) and Kurucz (1992) for \ion{Fe}{III}. 
The oscillator strengths for some lines of light elements which could not 
be found in the papers cited were supplemented from Kurucz' line list, from 
which also the damping constants for all metal lines were extracted.

Equivalent widths of metal lines between 5m\AA\ and 79m\AA\ were measured.
Since fewer metal lines are present in the spectra of KPD\,2109+4401 and 
PG\,1219+534 than in those of Feige\,48 and PG\,1605+072, we use the latter 
spectra as guidance to determine  
upper limits for metal abundances of the former from the absence of 
the presumedly strongest lines of the ions in question. 
An upper limit to their equivalent width of 5m\AA\ was assumed.     

\subsection{Feige\,48}

Since no He~II line can be detected, the helium ionization equilibrium can 
not be evaluated. 

\begin{figure*}
\vspace{11.0cm}
\includegraphics{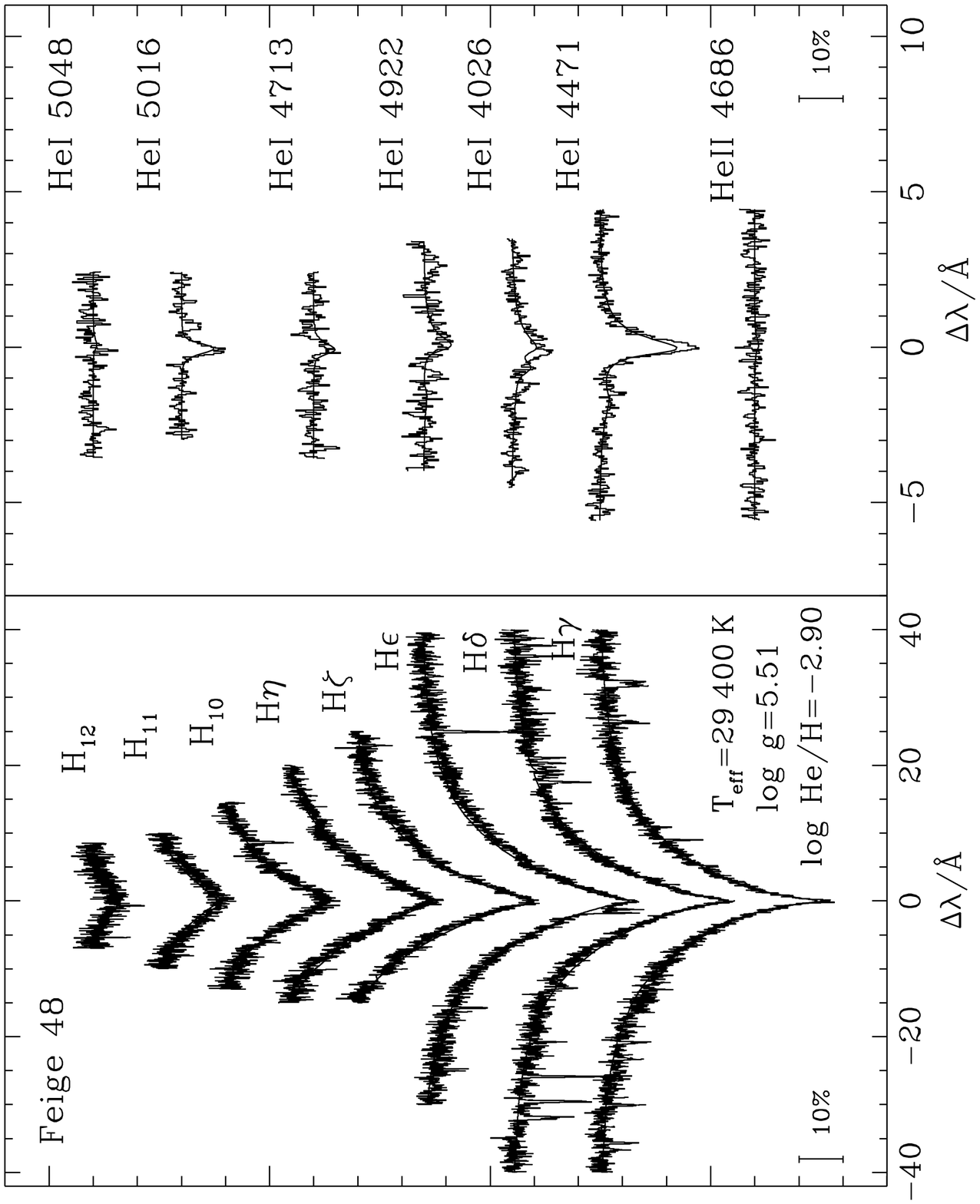}
\caption[]{Balmer and He line profile fits for Feige\,48 of the 
HIRES spectrum. \label{feige48fit}
}
\end{figure*}

The results of the analysis of the H and \ion{He}{I} lines for Feige 48
are listed in Table~\ref{res_f48} and compared to published values obtained 
from low resolution spectra. The formal errors of our fits 
are much smaller than the systematic errors (see below). The fit of the 
lines using the NLTE model grid is displayed in Fig.~\ref{feige48fit}. 
The agreement with the results from low resolution spectra analysed with 
similar LTE model atmospheres
(Koen et al. 1998) is
encouraging. 

Helium is
deficient by a factor of 80 and therefore
Feige$\,$48 has the lowest helium abundance among our 
programme stars.

\begin{table}
\caption{Atmospheric parameters for Feige\,48 from different methods, 
see text}\label{res_f48}
\begin{tabular}{|l|lll|}
\hline
method         & \Teff\ [K]      & log\,g        & \loghe \\
\hline
Koen et al. (1998)        & 28\,900$\pm$300 & 5.45$\pm$0.05 & --\\
LTE: H+He [M/H]=0.0       & 29\,500 & 5.53 & -2.94\\
LTE: H+He [M/H]=-2.0      & 30\,100 & 5.53 & -2.90\\
NLTE: H+He                & 29\,600 & 5.43 & -2.93\\
\hline
adopted              & 29\,500$\pm$300 & 5.5 $\pm$0.05 & -2.93$\pm$0.05\\
\hline
\end{tabular}
\end{table}
\begin{table}
\caption{Metal abundances for Feige\,48 compared to solar composition. n 
is the number of spectral lines per ion.}\label{abu_f48}
\begin{center}
\begin{tabular}{|l|rll|}
\hline
ion            & n     & \loge     & [M/H] \\
\hline
\ion{C}{III}          & 2   & 7.36$\pm$0.03  & $-$1.27\\
\ion{N}{II}           & 16    & 7.47$\pm$0.17& $-$0.48 \\
\ion{N}{III}          & 2    & 7.80$\pm$0.11 & $-$0.15\\
\ion{O}{II}           & 29   & 7.79$\pm$0.12 & $-$1.09\\
\ion{Ne}{II}          & 3    & 7.10$\pm$0.31 & $-$0.69\\
\ion{Mg}{II}          & 1    & 6.91          & $-$0.6\\
\ion{Al}{III}         & 2    & 5.50$\pm$0.18 & $-$0.89\\
\ion{Si}{III}         & 4    & 6.39$\pm$0.32 & $-$1.01 \\
\ion{Si}{IV}          & 1    & 6.17          & $-$1.07 \\
\ion{S}{III}          & 1    & 6.15          & $-$1.01 \\
\ion{Fe}{III}         & 26   & 7.55$\pm$0.19 & $+$0.13  \\
\hline
\end{tabular}
\end{center}
\end{table}

Three species are represented by two 
stages of ionization (\ion{C}{II} and 
\ion{C}{III}, \ion{N}{II} and \ion{N}{III}, \ion{Si}{III} and \ion{Si}{IV}). 
Since these line ratios 
are
very temperature sensitive at the temperatures in 
question, we alternatively can 
derive \Teff\ and abundances by matching these ionization equilibria.  
Gravity is derived 
subsequently from the Balmer lines by keeping \Teff\ and \loghe\ fixed. These
two steps are iterated until consistency is reached. 
\ion{C}{II} is represented by the 4267\AA\ line 
only, which is known to give notoriously too low carbon abundances. Indeed 
the carbon ionization equilibrium can not be matched at any reasonable 
\Teff. The ionization equilibrium of N indicates a slightly higher 
effective temperature of \Teff$\approx$30\,700\,K 
than derived from the Balmer and neutral helium lines using 
Saffer's procedure whereas the silicon ionization equilibrium indicates a 
slightly lower effective temperature of \Teff$\approx$28\,000 K.
However, in the case of the nitrogen ionization equilibrium,
\ion{N}{III} is represented by two weak 
lines only, and in the case of silicon \ion{Si}{IV} is represented by one 
weak line only. Considering the measurement errors for these lines, the 
nitrogen and silicon ionization equilibria can be considered to be in good
agreement with the fit of the Balmer and neutral helium lines.    

The effective temperature derived from the metal ionization equilibria is 
higher than that from Saffer's procedure. Whether this may 
be caused by an NLTE effect has to be investigated from detailed NLTE 
calculations for N and Si which are beyond the scope of this paper.

\begin{figure}
\vspace{8.0cm}
\includegraphics{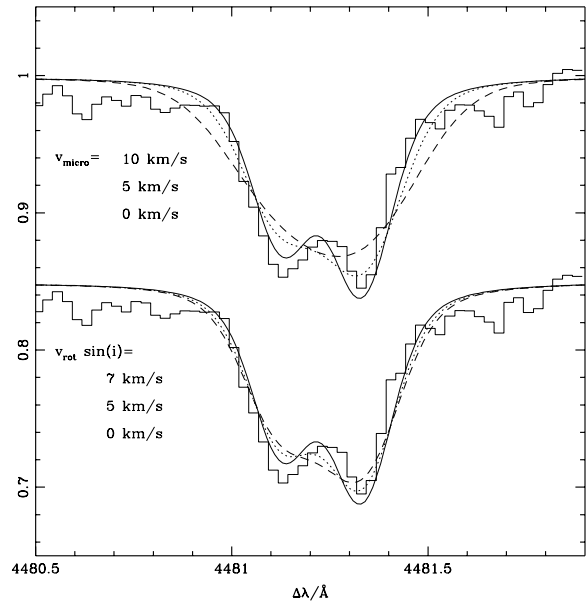}
\caption[]{Fit of the \ion{Mg}{II} doublet for microturbulent velocities (top) 
of v$_{micro}$=0, 5 (dotted), and 10\,km/s (dashed) and projected rotational 
velocities (bottom) of v$_{rot}$\,sin\,i=0, 5 (dotted), and 7 km/s (dashed) 
\label{feige48mg2}}
\end{figure}

Although several O lines are available in the spectrum of 
Feige$\,$48, it was impossible to determine the 
microturbulent velocity in the usual way, i.e. by removing the
slope of the O abundances with equivalent widths, due to the 
lack of sufficiently strong lines. 
However, the spectrum is very sharp lined and several close blends are 
resolved, e.g. the doublet structure of \ion{Mg}{II} 4481\AA\ is clearly 
seen (Fig.~\ref{feige48mg2}). The \ion{Mg}{II} doublet is best fitted with no 
extra broadening (see Fig.~\ref{feige48mg2}, top panel), 
i.e. v$_{micro}\le$3\,km/s. Alternatively, we can constrain the 
projected rotational velocity (see Fig.~\ref{feige48mg2}, bottom panel),  
an upper limit of v$_{rot}$\,sin\,i$\le$5\,km/s was adopted. 
Microturbulent and projected rotational velocities can also be constrained
from unblended lines such as \ion{Si}{III} 4552, 4567\AA\ 
although the limits are not as strict (v$_{micro}\le$5\,km/s, 
v$_{rot}$\,sin\,i$\le$10\,km/s) as those derived from the \ion{Mg}{II} doublet.
We adopt v$_{micro}$=0\,km/s. An increase to 5\,km/s translates 
into small systematic abundance uncertainties of 0.05dex for most 
ions. 
A temperature uncertainty of $\Delta$T$_{\rm eff}$=1000$\,$K translates 
into abundance uncertainties of less than 0.1$\,$dex. Hence systematic errors 
are smaller for most ions than the statistical errors. The resulting 
abundances are listed in Table~\ref{abu_f48} and plotted in 
Fig.~\ref{abundances}.

%

\subsection{KPD$\,$2109$+$4401}

\begin{table}
\caption{Atmospheric parameters for KPD\,2109+4401
from different methods, 
see text}\label{res_kpd}
\begin{tabular}{|l|lll|}
\hline
method         & \Teff\ [K]      & log\,g        & \loghe \\
\hline
Bill\`eres et al. (1998)& 31\,200 & 5.84  &   \\
LTE: H+He [M/H]=0.0       & 31\,200 & 5.77 & -2.22\\
LTE: H+He [M/H]=-2.0      & 31\,400 & 5.79 & -2.31\\
LTE: He                   & 32\,400 & 5.81 & -2.13\\
NLTE: H+He                & 31\,200 & 5.69 & -2.30\\
NLTE: He                  & 32\,500 & 5.73 & -2.18\\
\hline
adopted                   & 31\,800$\pm$600 & 5.76$\pm$0.05 & -2.23$\pm$0.1\\
\hline
\end{tabular}
\end{table}
\begin{table}
\vspace*{0.7cm}
\caption{Metal abundances for KPD\,2109+4401 compared to solar composition. n 
is the number of spectral lines per ion.}\label{abu_kpd}
\begin{center}
\begin{tabular}{|l|rll|}
\hline
ion            & n     & \loge     & [M/H] \\
\hline
\ion{C}{III}          & -    & $<$6.3  & $<-$2.3\\
\ion{N}{II}           & 2    & 7.26$\pm$0.30 & $-$0.69 \\
\ion{N}{III}          & 2    & 7.23$\pm$0.03 & $-$0.72\\
\ion{O}{II}           & -    & $<$6.6 & $<-$2.3\\
\ion{Ne}{II}          & -    & $<$6.5 & $<-$1.2\\
\ion{Mg}{II}          & -    & $<$5.9  & $<-$1.6\\
\ion{Al}{III}         & -    & $<$5.3  & $<-$1.1 \\
\ion{Si}{III}         & -    & $<$5.4  & $<-$2.1 \\
\ion{Si}{IV}          & -    & $<$5.4  & $<-$2.1 \\
\ion{S}{III}          & 2    & 6.50$\pm$0.02 & $-$0.66 \\
\ion{Fe}{III}         & 12    & 7.63$\pm$0.21 & +0.21  \\
\hline
\end{tabular}
\end{center}
\end{table}

\begin{figure*}
\vspace{11.0cm}
\includegraphics{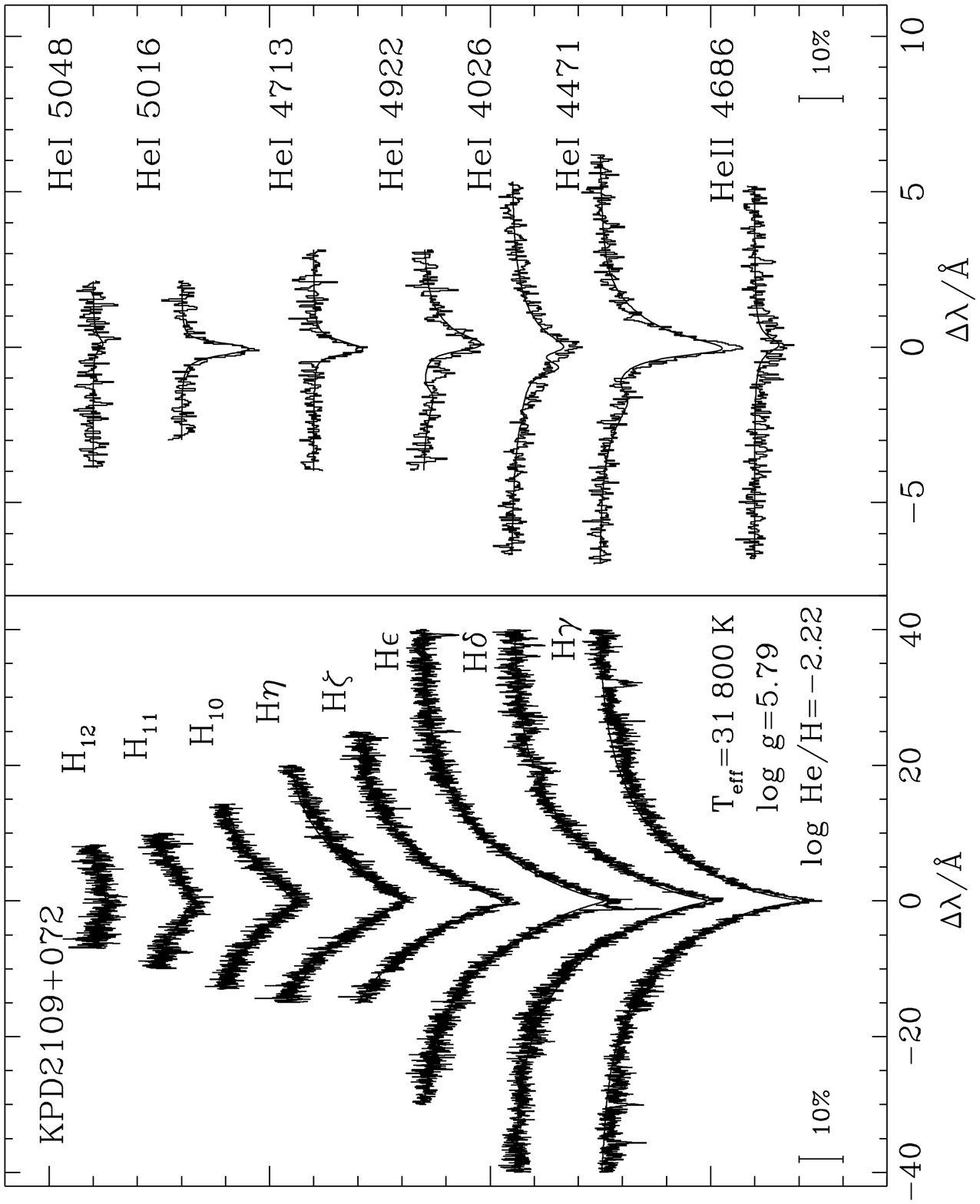}
\caption[]{Balmer and He line profile fits for KPD$\,$2109$+$4401 
of the HIRES spectrum.\label{kpdfit}
}
\end{figure*}

As in the case of PG\,1605$+$072 (paper I), 
\ion{He}{II} 4686\AA\ can be measured in the spectrum of 
KPD$\,$2109$+$4401 allowing the helium ionization equilibrium to be 
exploited as well as Saffer's procedure. Both methods 
give parameters in resonable agreement (see
Table~\ref{res_kpd}) and the averaged numbers 
(T$_{\rm eff}$=31$\,$800K, log~g=5.79 
log(He/H)$=-$2.22) were adopted for KPD$\,$2109$+$4401. The line profiles 
calculated from this model reproduces the observed spectrum well (see
Fig.\ref{kpdfit}). The derived \Teff\ and \logg\ are also in good agreement 
with the results from low resolution spectroscopy (Bill\`eres et al., 1998).
Nitrogen is also 
present in two stages of ionization (\ion{N}{II} and \ion{N}{III}) 
and the effective temperature derived 
from its ionization equilibrium is in perfect agreement with the results 
from Balmer and helium line fitting.

Helium is depleted by a factor of 17 with respect to the sun. 

From metal line profile fitting, as described for Feige\,48,
the projected rotational velocity of KPD$\,$2109$+$4401 is constrained by 
our spectra to v$\,$sini $\le$ 10$\,$km/s. The microturbulent 
velocity is constrained to v$_{micro}\le$5\,km/s.
Again  we adopted v$_{micro}$=0\,km/s for further analysis. 

Abundances of nitrogen, sulfur and iron were derived from the 
equivalent widths of their spectral lines. Upper limits for the abundances 
of carbon, oxygen, neon, magnesium, aluminium, and silicon were derived from 
the absence 
of the presumedly strongest lines of those ions.     

The resulting abundances are listed in Table~\ref{abu_kpd} and plotted in 
Fig.~\ref{abundances}.

Fontaine (1999, priv. comm.) obtained an improved light curve of 
KPD$\,$2109$+$4401 and detected some structure in the Fourier spectrum, in 
particular, a symmetric quintuplet, which may be due to 
rotational splitting.  A rotational period of about 6.6 days (or longer if  
not all the components of an n-tuplet are seen) has been inferred. 
Assuming a mass of 0.5\Msol\ we 
derive a radius of 0.16\Rsol\ from its gravity, which results in a 
rotational velocity of 1.2 km/s, too small to be 
detected in our HIRES spectra. 

\subsection{PG$\,$1219$+$534}

\begin{table}
\caption{Atmospheric parameters for PG\,1219+534
 from different methods, 
see text}\label{res_pg1219}
\begin{tabular}{|l|lll|}
\hline
method         & \Teff\ [K]      & log\,g        & \loghe \\
\hline
Koen et al. (1999)   & 32\,800$\pm$300 & 5.76$\pm$0.04 & --\\
LTE: H+He [M/H]=0.0       & 33\,200 & 5.93 & -1.60\\
LTE: H+He [M/H]=-2.0      & 33\,200 & 5.95 & -1.62\\
LTE: He                   & 35\,200 & 6.03 & -1.41\\
NLTE: H+He                & 33\,300 & 5.85 & -1.56\\
NLTE: He                  & 35\,400 & 5.87 & -1.38\\
\hline
adopted (see text)        & 34\,300$^{+2000}_{-1000}$ & 5.95$\pm$0.1 & -1.5$\pm$0.1\\
\hline
\end{tabular}
\end{table}
\begin{table}
\vspace*{0.7cm}
\caption{Metal abundances for PG\,1219+534 compared to solar composition. n 
is the number of spectral lines per ion.}\label{abu_pg1219}
\begin{center}
\begin{tabular}{|l|rll|}
\hline
ion            & n     & \loge     & [M/H] \\
\hline
\ion{C}{III}          & -    & $<$6.6           & $<-$2.0  \\
\ion{N}{II}           & 7    & 7.79$\pm$0.08    & $-$0.16  \\
\ion{N}{III}          & 5    & 7.79$\pm$0.16    & $-$0.16\\
\ion{O}{II}           & -    & $<$7.0           & $<-$1.9  \\
\ion{Ne}{II}          & -    & $<$6.7           & $<-$1.0  \\
\ion{Mg}{II}          & -    & $<$6.2           & $<-$1.3  \\
\ion{Al}{III}         & -    & $<$5.7           & $<-$0.7  \\
\ion{Si}{IV}          & -    & $<$5.6           & $<-$1.9  \\
\ion{S}{III}          & 2    & 6.86$\pm$0.06    & $-$0.30 \\
\ion{Fe}{III}         & 2    & 7.56$\pm$0.16    & +0.13  \\
\hline
\end{tabular}
\end{center}
\end{table}
 
\begin{figure*}
\vspace{22.0cm}
\includegraphics{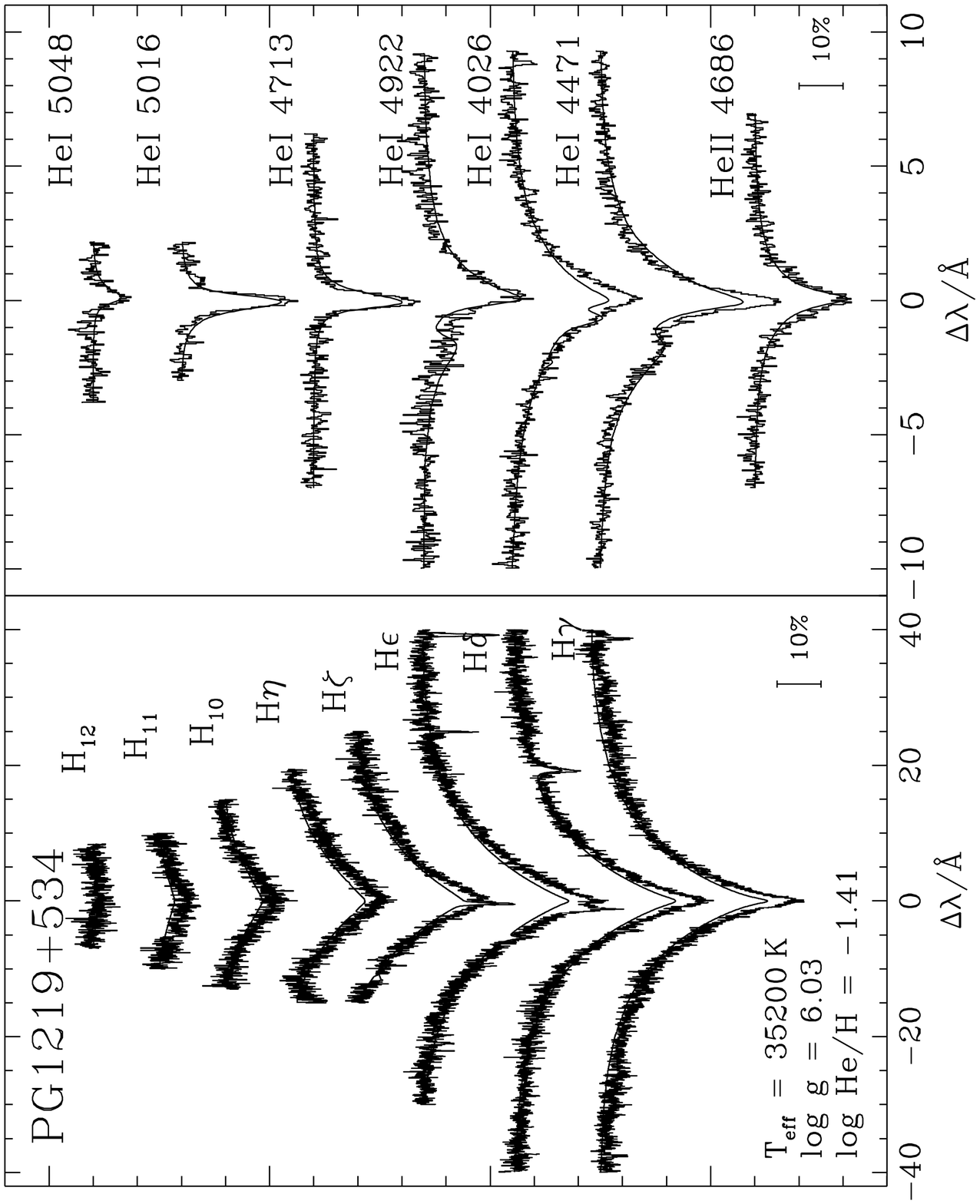}
\includegraphics{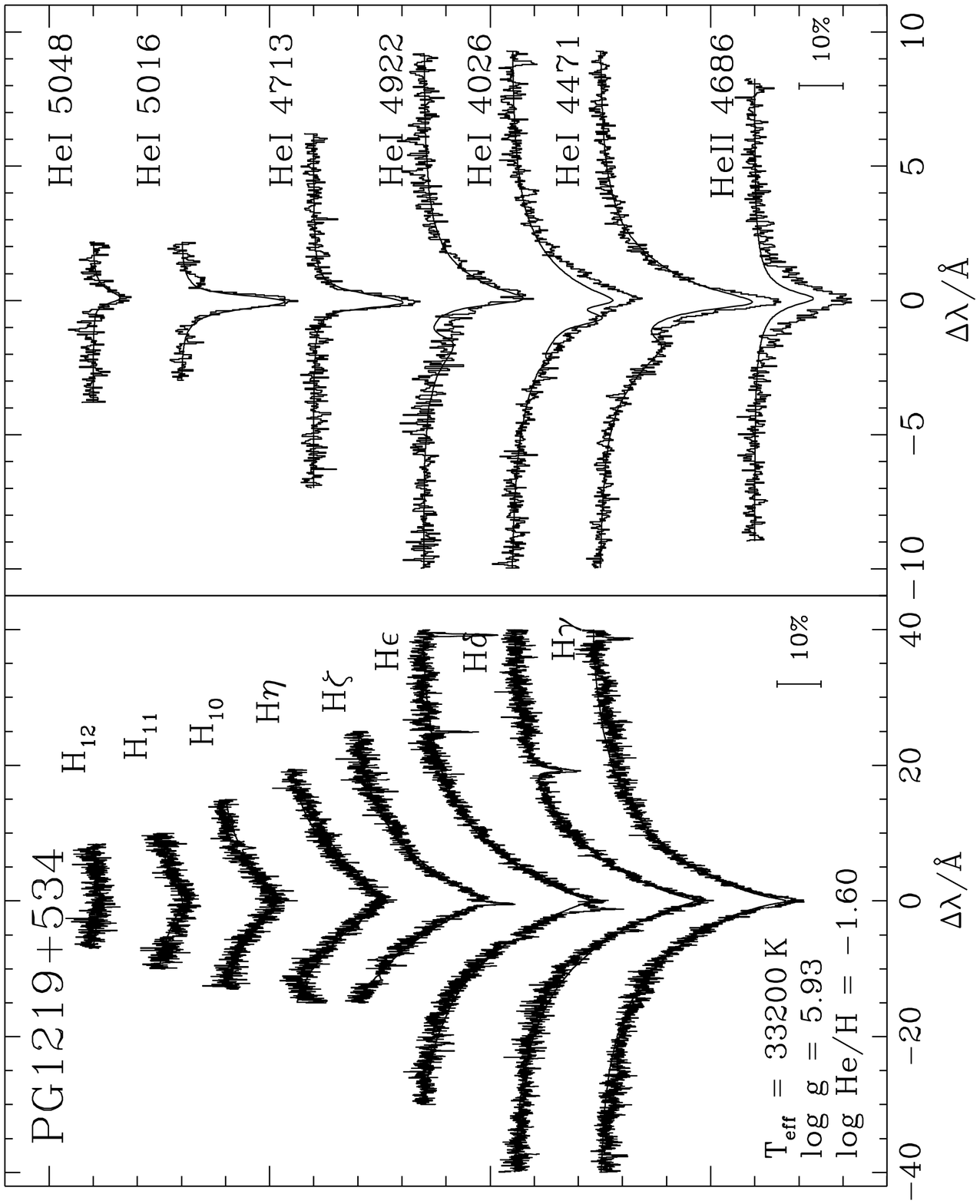}
\caption[]{Top: Balmer and He line profile fits for PG$\,$1219$+$534 of the HIRES 
spectrum. 
Note the mismatch of the He~II 4686\AA\ line profile and the cores of 
He~I 4026\AA\ and 4471\AA.\\
Bottom: He line profile fits for PG$\,$1219$+$534 of the HIRES spectrum
to determine T$_{\rm eff}$ and log(He/H) 
simultaneously, log~g is adjusted to match the Balmer line 
wings. Note the mismatch of the cores of the Balmer lines and of He~I 4026\AA\ 
and 4471\AA.\label{pg1219fit}}
\end{figure*}

As already pointed out by Koen et al. (1999) the helium lines are stronger 
than in most other sdB stars. \ion{He}{II} 4686\AA\ is detected allowing 
the He ionization equilibrium to be used as a temperature indicator. 
\Teff\ and \logg\ derived with Saffer's procedure agree well with 
results from low resolution spectroscopy (Koen et al., 1999). However, unlike 
PG\,1605$+$072 and KPD\,2109$+$4401, the  helium ionization 
equilibrium and 
Saffer's procedure yield discrepant results (see Table~\ref{res_pg1219}):
T$_{\rm eff}$=33$\,$200K, 
log~g=5.93, log(He/H)=-1.60 (Saffer's  procedure, Fig.~\ref{pg1219fit}, 
top panel) and 
T$_{\rm eff}$=35$\,$200K, log~g=6.03, log(He/H)=-1.41 (He ionization 
equilibrium, Fig.~\ref{pg1219fit}, bottom panel). 
At the lower T$_{\rm eff}$ the Balmer 
lines are well 
matched throughout the entire profile, whereas for He~II 4686\AA\ there is 
a significant mismatch (see Fig.~\ref{pg1219fit}, top panel). 
At the higher T$_{\rm eff}$ He~II 4686\AA\
is well reproduced, but the Balmer line cores are not reproduced at all 
(see Fig.~\ref{pg1219fit}, bottom panel). 
The line cores of He~I 4026\AA\ and 4471\AA\ cannot be reproduced by 
either model. 
Despite of its high gravity PG$\,$1219$+$534 has an unusually 
high helium abundance, i.e. helium is deficient by a factor of 2 to 5, only.


Besides the Balmer and the helium lines, metal lines of \ion{N}{II}, 
\ion{N}{III}, \ion{S}{III} and \ion{Fe}{III} are present. 
The ionization equilibrium of nitrogen (7 \ion{N}{II} and 5 \ion{N}{III} lines)
can be used to determine the effective temperature, which requires 
the effective temperature to be as high as \Teff=36800K.
At the low \Teff\ indicated by the Balmer line fit, the nitrogen abundance 
derived from \ion{N}{II} lines differs from that derived from \ion{N}{III} 
lines by almost 1 dex.
 
The abundances were calculated from the model most consistent with the 
nitrogen ionization equilibrium and results are listed in 
Table~\ref{abu_pg1219} and plotted in Fig.~\ref{abundances}. The 
uncertainty in \Teff\ was taken into account when we estimated the upper 
limits for carbon, oxygen, neon, magnesium, aluminium and silicon from the 
absence of their presumedly strongest lines.

Since the metal lines are very sharp we can constrain 
either the microturbulent velocity or the projected rotational velocity as 
demonstrated for Feige\,48. 
Due to the weakness of the lines the limits are not as tight as for Feige\,48:
v$_{micro}\le$5\,km/s, v$\,$sini $\le$ 10$\,$km/s.

\section{Discussion} 

We have determined atmospheric parameters (\Teff, \logg, He/H), metal 
abundances and projected rotation velocities from time averaged, 
high resolution spectra for four pulsating sdB stars (Paper I, this paper). 
The effective temperatures and gravities (see Fig.~1) 
confirm that KPD\,2109+4401, 
PG\,1219+534 and Feige\,48 are bona-fide extended horizontal branch stars, 
the latter being on the terminal EHB line, whereas the gravity of PG\,1605+072
is too low to be consistent with an EHB nature and might indicate a post-
EHB evolutionary status for this star.

\subsection{Atmospheric parameters}

Using line blanketed LTE model atmospheres for solar and metal poor 
composition ([M/H]=-2.0) as well as line blanketed zero metalicity NLTE model
atmospheres we found that 
\begin{itemize}
\item metalicity effects are small.  
\item NLTE effects are unimportant for the temperature determination.
\item there is an offset between gravities determined from the 
NLTE grid and the LTE grid. The NLTE gravities are slightly lower, by up 
to 0.1\,dex.
\end{itemize}

A careful investigation of several temperature indicators, i.e. line profile 
fitting using Saffer's procedure, the ionization equilibria of helium, 
nitrogen and silicon gave consistent results for Feige\,48, KPD2109$+$4401
and PG\,1605$+$072 (paper I). However, for PG\,1219+534 a considerably higher 
effective temperature was derived from the ionization equilibria of 
nitrogen (36\,800K) and helium (34\,400\,K) than from the line profile 
fitting (Saffer's method, 33\,200\,K). What causes these discrepant results
? NLTE effects might be considered. However, our NLTE calculations show 
that this is not the case for helium. For nitrogen such calculations are 
beyond the scope of this paper. In view of the consistent results for the 
other stars, NLTE effects for the N ionization equilibrium in PG\,1219+534 
appear unlikely. 

PG\,1219+534 has the highest He abundance and the shortest pulsation period 
of our programme stars. Hence the discrepancy could be related to these 
properties. Since the helium lines are stronger than in the other programme 
stars, line broadening is more important for the He~I lines in the spectrum of 
PG\,1219+534. We used tabulations of Barnard et al. (1969), Shamey (1969), 
Barnard et al. (1974, 1975) and Griem (1974) to compute 
the He~I line profiles. New calculations have become available recently 
(Beauchamp et al., 1997). However, these tables are tailored for the use in 
white dwarf atmospheres and cannot be used for sdB stars because they do not 
extend to the significantly lower densities in the outer layers of sdB 
atmospheres. An extension of Beauchamp et al.'s  tables 
would be useful for the sdB star line profile synthesis. For \ion{He}{II} 
lines we used the tabulations of Sch\"oning \& Butler (1989). 

Due to diffusion the helium distribution in the atmospheres could be 
inhomogeneous. 
If helium accumulates in the deeper atmospheric layers, the predicted 
\ion{He}{I} line profiles, however, would be even broader and shallower, 
increasing the discrepancy with observations. A vertical stratification of 
helium, therefore, appears to be unlikely.

On the other hand the short pulsational periods 
(P=128.1s$\ldots$148.8s) observed 
for PG\,1219+534 might indicate that the model assumption of hydrostatic 
equilibrium does no longer hold for the outer layers of the atmosphere, 
where the cores of the Balmer and \ion{He}{I}, 4026, 4471\AA\ lines are 
formed. In this case \Teff\ derived from the He and N ionization equilibria 
should be preferred since the relevant lines form in deeper atmospheric 
layers.

Since the origin for the discrepancy remains unclear we adopted weighted 
means for the atmospheric parameters and had to admit a larger error range 
for PG\,1219+534 than for the other programme stars.

\subsection{Abundances}

\begin{figure*}
\vspace{15.0cm}
\includegraphics{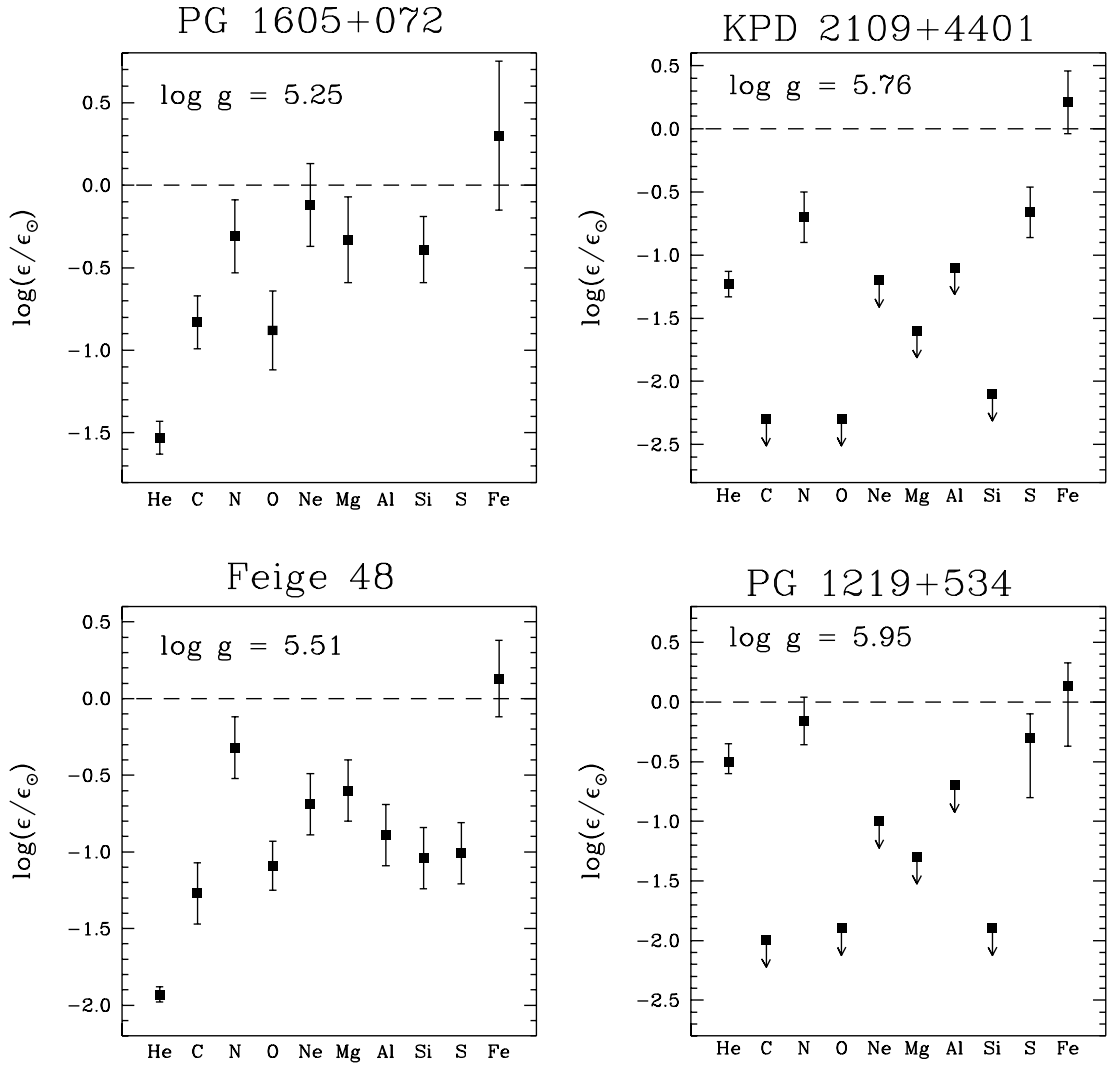}
\caption[]
{Abundances of the programme stars relative to solar values\label{abundances}}
\end{figure*}

The resulting abundances are plotted in Fig.~\ref{abundances} and compared 
to PG\,1605+072 (paper I). Although KPD\,2109$+$4401 and PG\,1219$+$534 
do not display C, O, Ne, Mg Al and Si lines, stringent  
upper limits have been derived from the absence of the presumedly strongest 
lines of these species.

Helium is depleted in all programme stars. Its abundances ranges from 
1/80 solar (Feige\,48) to 1/3 solar (PG\,1219+534). 

From the interplay of gravitational settling and radiative levitation, an
anticorrelation between helium abundance and gravity might be suspected. 
However no such correlation exists. This has also been found previously from 
studies of much larger samples of non-variable sdB stars 
(e.g. Schulz et al., 1991, Fontaine \& Chayer, 1997).

Like helium the metals are deficient with respect to the sun. Iron, however, 
is a notable exception, 
since it is solar to within the error limits in all programme stars. 

Large deficiencies of carbon, oxygen and silicon of up to 2 dex or 
more (KPD$\,$2109$+$4401) 
have been found, whereas nitrogen and sulfur are only mildly deficient. 
In PG\,1219+534 nitrogen is even almost solar. Large deficiencies of carbon 
and silicon as well as near solar 
nitrogen abundances have been reported for several non-variable sdB stars 
from high resolution UV spectroscopy (e.g. Heber et al. 1984b, Lamontagne et al.
1985, 1987, see Heber, 1998 for a review). 

It is worthwhile to note that the abundances of 
C, O, Ne, Mg, Al and Si in the high gravity programme stars 
KPD\,2109+4401 and PG\,1219+534 are considerably lower than in the 
lower gravity stars PG\,1605+072 and Feige\,48 
(see Fig. \ref{abundances})
which may point to the (selective) action of diffusion,
i.e. an interplay between gravitational settling 
and radiative levitation. 
It is, however, puzzling that iron is 
solar irrespective of the stellar gravity. 

Recently, Ohl et al. (2000) analysed the far UV spectrum of PG\,0749+658, 
a sdB star somewhat cooler than our 
programme stars, obtained with the FUSE satellite and found solar abundances 
for several iron group elements.
PG\,0749+658 shares the mild depletion of nitrogen and sulfur with our 
programme stars.
KPD\,2109+4401 and PG\,1219+4401 show a similarly strong deficiency of 
carbon and silicon as PG\,0749+658 and several other non-variable sdBs 
(Baschek et al., 1982a,b; Lamontagne et al., 1985, 1987, Heber et al. 1984b). 

However, until recently only a few diffusion 
calculations were available (e.g. Michaud et al. 1985). Photometric and 
spectroscopic observations of blue horizontal branch stars in 
globular clusters (Grudahl et al., 1999, Moehler et al., 1999, 
Behr et al., 1999a)
stimulated calculations of atmospheric diffusion in atmospheres for such 
stars (Hui-Bon-Hoa et al., 2000), which are, however, 
considerably cooler than our programme stars. The predicted iron abundances 
are even larger than the observed ones.  

For the sdB star PG\,0749+658, Ohl et al. (2000)
calculated equilibrium abundances 
within the framework of the radiative levitation theory and were able to 
reproduce the observed abundances of carbon, sulfur and iron, but predicted 
higher abundances  for nitrogen and silicon and a much lower 
helium abundance than observed. 
Mass loss has frequently been invoked to explain the 
discrepancy between observed and predicted helium abundances (Fontaine \& 
Chayer, 1997). 
Unglaub \& Bues (1998) carried out diffusion calculation for 
helium, carbon and 
nitrogen for a model of \Teff=40\,000\,K, \logg=6.0 including mass loss 
processes. For mass loss rates less than 10$^{-13}$ \Msol/yr 
the predicted helium, carbon and nitrogen abundance are 
qualitatively consistent with the
observation. For a quantitative comparison additional calculations 
at lower effective temperatures and gravities as well as for various mass
loss rates are required.  
Finally, we point out that a solar surface iron  
abundance is consistent with the diffusion calculations of 
Charpinet et al. (1997) for the envelopes of pulsating sdB star models.

\subsection{Rotation velocities}

The spectral lines of PG$\,$1605$+$072 are considerably broadened,
which Heber et al. (1999) attributed to stellar rotation and derived
v$\,$sin$\,$i = 39km/s, by fitting the strongest metal lines. Recently,
O'Toole et al. (2000) performed time-series spectroscopy of PG$\,$1605$+$072
and detected radial velocity variations at the same frequencies found from 
photometry with amplitudes up to 14 km/s at H$\beta$. 
Somewhat smaller 
velocity amplitudes were found for higher Balmer lines. 
The authors conclude 
that these velocity 
variations arise from the pulsations. Since part of the 
line broadening observed in the time averaged spectrum analysed in paper I 
may be due to radial velocity variations, the true projected rotational 
velocity is slightly smaller than v$\,$sin$\,$i = 39km/s. 
Feige$\,$48, PG$\,$1219$+$534 and KPD$\,$2109$+$4401 
are all very sharp-lined and we derived upper limits 
of v$\,$sin$\,$i$<$5--10km/s (see above). This is consistent with 
time-series spectroscopy of KPD\,2109+4401 (Jeffery \& Pollaco, 2000) who 
measured radial velocity variations of 2\,km/s, well below our detection 
limit. The same holds for any rotational broadening  that has been inferred
from the analysis of the light curve (Fontaine, 1999, priv. com., 
v$_{rot}$\,sin\,i=1.2 \,km/s or less).

The low projected rotational velocities of the pulsating sdB stars 
may be compared to the rotation velocities found for globular cluster 
horizontal branch stars, since sdB stars form the extremely hot end of the 
horizontal branch. Behr et al. (2000a,b) analysed Keck HIRES spectra
of HB stars in the globular clusters M\,13 and M\,15 and found that 
a star's rotation appears to be related to its position along the horizontal 
branch. 
Stars hotter than about 11\,000\,K  were found to be rotating slowly 
(v$\,$sin$\,$i $<$ 10km/s) whereas the cooler horizontal branch stars 
rotate more rapidly (v$\,$sin$\,$i $\approx$ 40km/s for M\,13) as 
previously observed by Peterson et al. (1995) for M\,13. The low rotational 
velocities of our extreme horizontal branch stars are therefore consistent 
with the slow rotation derived for the hottest 
stars in M\,13 and M\,15. The rotating PG\,1605+072 hence appears to be an 
exceptional case. 

SdB stars are generally believed to evolve directly into white dwarfs which 
are mostly slow rotators as well (Heber et al., 1997, Koester et al, 1998).   

\section{Appendix: Interstellar \ion{Ca}{II} K lines}

Radial velocities and equivalent widths of interstellar 
\ion{Ca}{II} K are derived by fitting the observed profile with three 
(PG\,1219+534) or four Gaussian profiles. Results are listed in Table~\ref{IS}.

\begin{table}
\caption{Radial velocities and equivalent widths of interstellar 
\ion{Ca}{II} K derived by fitting the observed profile with three 
(PG\,1219+534) or four Gaussian profiles.}\label{IS}
   \begin{center}
      \begin{tabular}{|r r| r r| r r| r r|}
\hline
\multicolumn{2}{|c}{Feige\,48}& \multicolumn{2}{|c}{KPD\,2109+4401} &
\multicolumn{2}{|c}{PG\,1219+534} & \multicolumn{2}{|c|}{PG\,1605+072}\\
\hline
v & $W_{\lambda}$& v & $W_{\lambda}$ & v  & $W_{\lambda}$ & v & $W_{\lambda}$\\
km/s & m\AA\  &km/s & m\AA\    & km/s & m\AA\    & km/s 
& m\AA\  \\
\hline
$-$55  &  63   &  $-$42   &  41  & $-$57   &  78  & $-$55   &   8  \\
$-$38  &  38   &  $-$13   &  86  & $-$28   &  70  & $-$39   &  14  \\
$-$17  &  53   &      3   &  31  & $-$5    &  26  & $-$18   &  96  \\
  5    &  12   &     11   &  11  &         &      &     0   &  68  \\
\hline
\end{tabular}
\end{center}
\end{table}

\begin{acknowledgements}
We thank Ralf Napiwotzki for many fruitful discussions.
A travel grant to the W.M.Keck observatory by the Deutsche 
Forschungsgemeinschaft is
gratefully acknowledged.   

\end{acknowledgements}

\end{document}